\documentclass[pra,preprint,amssymb,amsmath,showpacs,nofootinbib]{revtex4}
\usepackage{amsthm}

\usepackage{latexsym}

\newtheorem{theorem}{Theorem}

\newtheorem{corollary}[theorem]{Corollary}
\newtheorem{proposition}[theorem]{Proposition}

\theoremstyle{remark}
\newtheorem*{claimnn}{Claim}
\newtheorem*{subclaimnn}{Subclaim}

\DeclareMathOperator{\supp}{supp}

\newcommand{\myqed}{\hfill \mbox{\raggedright \rule{0.45em}{1.4ex}}}

\newcommand{\myreal}{\mathbb{R}}

\newcommand{\ignore}[1]{}

\begin{document}


\title{Inequalities that Collectively Completely Characterize\\the Catalytic Majorization Relation}

\date{September 23, 2007}
\author{Matthew Klimesh}
\affiliation{Jet Propulsion Laboratory, California Institute of Technology, Pasadena, CA 91109}
\email{matthew.a.klimesh@jpl.nasa.gov}





\newcommand{\myhalf}{\mbox{$\frac{1}{2}$}}

\begin{abstract}
For probability vectors $x$ and $y$, the catalytic majorization relation $x \prec_T y$ is defined to hold when there exists a probability vector $z$ such that $x \otimes z$ is majorized by $y \otimes z$.  In this paper, an infinite family of functions is given such that, subject to some trivial restrictions, $x \prec_T y$ if and only if $f_r(x)<f_r(y)$ for all functions $f_r$ in the family.  An outline of a proof of this result is provided.  The catalytic majorization relation is known to provide a determination of which transformations of jointly held pure quantum states are possible using local operations and classical communication when an additional jointly held state may be specified to facilitate the transformation without being consumed.
\end{abstract}

\maketitle

\section{Introduction}  \label{sec:intro}
Let $x=(x_1,\ldots,x_d)$ and $y=(y_1,\ldots,y_d)$ be
$d$-dimensional vectors with real components.
Let $x^{\downarrow}$ denote the vector obtained by arranging
the components of $x$ in decreasing order:
$x^{\downarrow} = (x_1^{\downarrow},\ldots,x_d^{\downarrow})$
where $x_1^{\downarrow} \geq \cdots \geq x_d^{\downarrow}$.
Then $x$ is said to be \emph{majorized} by $y$, written $x \prec y$, if

\[ \sum_{i=1}^k x_i^{\downarrow} \leq
   \sum_{i=1}^k y_i^{\downarrow} \]
when $1 \leq k < d$ and $\sum_{i=1}^d x_i = \sum_{i=1}^d y_i$.  The majorization relation has been well studied; useful literature on the subject includes \cite{Mar79} and \cite{Bha97}.
%

For a $d$-dimensional vector $x$ and an $\ell$-dimensional vector $z$ the notation $x \otimes z$ denotes the $d\ell$-dimensional vector tensor product,
which is a vector whose components are all of the terms of the form $x_i z_j$.  For our purposes, the order of the components is irrelevant.
Our interest is in the following question: given two $d$-dimensional probability vectors $x$ and $y$, does there exist a (finite dimensional) probability vector $z$ such that $x \otimes z \prec y \otimes z$?  This question motivates the definition of the \emph{catalytic majorization} relation: we say $x$ is \emph{catalytically majorized} by $y$, written $x \prec_T y$, when such a $z$ exists.\footnote{The catalytic majorization relation is called the trumping relation in~\protect\cite{Daf01,Daf04,Nie99a}.}  Probability vectors of different dimensions can be compared with this relation by appending zeros to the shorter vector, and in fact we find it convenient to consider catalytic majorization as a relation among probability vectors of all finite dimensions.



In this paper, an infinite family of functions is given with the property that, subject to some trivial restrictions, $x \prec_T y$ if and only if $f_r(x)<f_r(y)$ for all functions $f_r$ in the family.  This result provides an answer to open problem~4 of \cite{Kru05} (also posed in \cite{Nie99a}), which asks to determine relatively simple conditions to decide whether or not $x \prec_T y$.  This result also essentially resolves a conjecture of Nielsen (as stated in \cite{Daf04}) on that problem; with some minor adjustments, that conjecture would hypothesize the present main theorem.

We regard the main mathematical significance of our result to be in clarifying the nature of the catalytic majorization relation, as well as providing a formulation that may be more useful mathematically than the definition.  However, we also note that our result should provide a means of determining whether $x \prec_T y$ for a given $x$ and $y$; the definition does not suggest a general practical method for making this determination.

The catalytic majorization relation has its origins in the field of quantum information.  In recent years in this field has developed rapidly
and much research has been directed toward understanding
what sort of manipulations of quantum-mechanically entangled states
are possible.  This research has in part been motivated by applications
of entanglement including quantum teleportation, quantum dense coding,
quantum cryptography, and quantum computation.

Both majorization and catalytic majorization have been
shown to arise in the study of transformations of entangled
bipartite pure quantum states.  Nielsen obtained the following result \cite{Nie99b}:
\begin{theorem}  \label{thm:nielsen}
 Suppose Alice and Bob are in joint possession of an entangled
 pure quantum state $|\psi_1\rangle$ that they wish to transform
 into another bipartite entangled pure state $|\psi_2\rangle$.
 Let $|\psi_1\rangle = \sum_{i=1}^d \sqrt{\alpha_i}|i_A\rangle |i_B\rangle$
 be a Schmidt decomposition of $|\psi_1\rangle$ and let
 $|\psi_2\rangle = \sum_{i=1}^d \sqrt{\beta_i}|i'_A\rangle |i'_B\rangle$
 be a Schmidt decomposition of $|\psi_2\rangle$.  Then $|\psi_1\rangle$
 can be converted to $|\psi_2\rangle$ (with success guaranteed) using
 only local operations
 and classical communication (LOCC) if and only if the vector
 $\alpha = (\alpha_1,\ldots,\alpha_d)$ is majorized by
 $\beta = (\beta_1,\ldots,\beta_d)$.
\end{theorem}

Jonathan and Plenio \cite{Jon99} extended this result by showing
that even if it is not possible to convert $|\psi_1\rangle$
to $|\psi_2\rangle$ directly (according to Theorem~\ref{thm:nielsen}),
it may be possible to convert $|\psi_1\rangle |\phi\rangle$
to $|\psi_2\rangle |\phi\rangle$, where $|\phi\rangle$ is an
additional bipartite state shared by Alice and Bob.  If $x$, $y$, and $z$
are the vectors of (squared) Schmidt coefficients of $|\psi_1\rangle$,
$|\psi_2\rangle$, and $|\phi\rangle$ respectively, then the
Schmidt coefficients of $|\psi_1\rangle |\phi\rangle$ are
the components of $x \otimes z$ and the Schmidt coefficients of
$|\psi_2\rangle |\phi\rangle$ are the components of $y \otimes z$;
thus Nielsen's Theorem implies that $|\psi_1\rangle |\phi\rangle$
can be converted to $|\psi_2\rangle |\phi\rangle$ when
$x \otimes z \prec y \otimes z$.  The state $|\phi\rangle$
is not consumed by this transformation; for this reason $|\phi\rangle$
is referred to as a catalyst and we say $|\phi\rangle$ catalyzes
the transformation from $|\psi_1\rangle$ to $|\psi_2\rangle$.
A state exists that can catalyze the transformation from
$|\psi_1\rangle$ to $|\psi_2\rangle$ if and only if $x \prec_T y$.


\section{Main Result}
To state our main theorem we specify a family of functions, indexed by a real number $r$.
For a $d$-dimensional probability vector $x$, let
\[
  f_r(x) =
  \begin{cases}
    \ln \sum_{i=1}^d x_i^r & (r>1); \\
    \sum_{i=1}^d x_i \ln x_i & (r = 1); \\
    -\ln \sum_{i=1}^d x_i^r & (0<r<1); \\
    -\sum_{i=1}^d \ln x_i & (r = 0); \\
    \ln \sum_{i=1}^d x_i^r & (r<0).
  \end{cases}
\]
If any of the components of $x$ are
$0$, we take $f_r(x) = \infty$ for $r \leq 0$.

Our main theorem is the following:
\begin{theorem}  \label{thm:main}
 Let $x=(x_1,\ldots,x_d)$ and $y=(y_1,\ldots,y_d)$ be $d$-dimensional
 probability vectors.
 Suppose that $x$ and $y$ do not both contain components equal to $0$
 and that $x \neq y$.  Then $x \prec_T y$ if and only if
 $f_r(x) < f_r(y)$ for all $r \in \myreal$.
\end{theorem}

We refer to the inequalities $f_r(x) < f_r(y)$ as the $f_r$ inequalities.  Note that the restrictions on $x$ and $y$ in Theorem~\ref{thm:main} do not limit the scope of the theorem in any essential way; for example, it is clear that adding or removing components that are $0$ from $x$ and $y$ does not affect the status of the catalytic majorization relation between them.

We previously reported the statement of Theorem~\ref{thm:main} in \cite{Kli04}.

It should be clear that there are many equivalent choices for the $f_r$ functions, since for any particular $r$ the composition of any increasing function with $f_r$ will give the same inequality in Theorem~\ref{thm:main}.  For example, $\sum_{i=1}^d x_i^r$ could be used in place of $\ln \sum_{i=1}^d x_i^r$ when $r>1$.  With our definition, for $r>0$ the function $f_r$ is, up to a positive constant factor that depends on $r$, the negative of the R\'{e}nyi entropy $H_r$.  In particular, the function $f_1$ is the negative of the Shannon entropy.

\section{Related Results in the Literature}
There are many related results in the literature; we attempt to list the most relevant here.

Recently, Aubrun and Nechita \cite{Aub07} obtained a result that has many aspects of our Theorem~\ref{thm:main}.  To describe their result, it is convenient to consider catalytic majorization to be a relation among infinite-dimensional probability vectors with a finite number of nonzero components (i.e., with finite support).  Aubrun and Nechita considered the set $T_c(y)$, defined to be the set of all $x$ such that $x \prec_T y$.  In terms of the $f_r$ functions, their main result implies that $x \in \overline{T_c(y)}$ if and only if $f_r(x) \leq f_r(y)$ for $r > 1$, where $\overline{T_c(y)}$ is defined to be the $\ell_1$ closure of $T_c(y)$ within the space of infinite probability vectors with finite support.  Stated another way, their result is that if $x$ and $y$ are infinite probability vectors with finite support, then the following are equivalent:
%
%
\begin{itemize}
\item[(i)] For any $\epsilon>0$ there exists an $x'$ with finite support
such that $\|x-x'\|_1 < \epsilon$ and $x' \prec_T y$.
\item[(ii)] The inequalities $f_r(x) \leq f_r(y)$ hold for $r > 1$.
\end{itemize}

We point out a subtlety that arises when comparing this result to ours.  Let $P_d$ be the set of all infinite probability vectors with at most $d$ nonzero components, and let $T(y) = \{ x \in P_d : x \prec_T y \}$.  Then, as noted in \cite{Aub07}, even though $T(y) = T_c(y) \cap P_d$, the $\ell_1$ closure $\overline{T(y)}$ is generally a strict subset of $\overline{T_c(y)} \cap P_d$.  Essentially, for some $d$-dimensional vectors $x$, one cannot find an $x'$ that is ``close'' to $x$ for which $x' \prec_T y$ unless $x'$ can have more nonzero components than $x$.  Alternatively, this can be regarded as an indication of the importance of the $f_r$ inequalities with $r<1$ for characterizing $T(y)$.



The result of Aubrun and Nechita that we have stated above can be obtained as a consequence of our Theorem~\ref{thm:main}.  However, their main result also applies to multiple-copy transformations, which are not addressed by the present work.

Transformations of states that use catalyst states as described in Section~\ref{sec:intro} are said to be transformations that use entanglement-assisted LOCC (ELOCC)\@.  Thus Theorem~\ref{thm:main} can be regarded as a characterization of transformations that are possible under ELOCC\@.

If for some positive integer $k$, the state $|\psi_1\rangle^{\otimes k}$ (i.e., a state that is $k$ copies of $|\psi_1\rangle$) can be transformed to $|\psi_2\rangle^{\otimes k}$ under LOCC, then we say that the state $|\psi_1\rangle$ can be transformed to $|\psi_2\rangle$ under multiple-copy LOCC (MLOCC), and the transformation is called a multiple-copy transformation.  It was observed in \cite{Ban02} that MLOCC allows transformations that are not possible under LOCC\@.  Duan et al.\ \cite{Dua05} showed that any transformation that is possible under MLOCC is also possible under ELOCC\@.  Additional results on MLOCC and its close relation to ELOCC can be found in \cite{Fen06,Dua05}.  The main result of Aubrun and Nechita \cite{Aub07} implies a characterization of MLOCC that is identical to their characterization of ELOCC\@.

Vidal \cite{Vid99} showed that Nielsen's theorem (\cite{Nie99b}, stated above as Theorem~\ref{thm:nielsen}) can be generalized to provide the optimum probability of transforming a state $|\psi_1\rangle$ to another state $|\psi_2\rangle$ under LOCC\@.  (Nielsen's theorem covers the case where the transformation can be made with certainty.)  It was noted by Jonathan and Plenio \cite{Jon99} that catalyst states can be useful in probabilistic entanglement transformations.  Properties of catalyst-assisted probabilistic entanglement transformations were investigated in \cite{Daf04,Fen05}.

We do not consider MLOCC or probabilistic entanglement transformations in this paper.








\section{Additive Schur-Convex Functions}  \label{sec:asc}
A useful tool in the study of majorization is the notion of
a \emph{Schur-convex} function.  A function $f: \myreal^d \rightarrow \myreal$
is Schur-convex if
$f(x) \leq f(y)$ whenever $x \prec y$ \cite{Mar79}.  
Nielsen \cite{Nie99a} has introduced the notion of an \emph{additive Schur-convex} function:
A function $f$ from
probability vectors (of any dimension) to $\myreal$ is additive Schur-convex if $f$ is Schur-convex (when restricted
to probability vectors of a given dimension), and
$f(x \otimes y) = f(x) + f(y)$.  If $x \prec_T y$
then we must have $f(x) \leq f(y)$ for any additive Schur-convex
function $f$, since if $x \otimes z \prec y \otimes z$ then
$f(x)+f(z) = f(x \otimes z) \leq f(y \otimes z) = f(y)+f(z)$.
Clearly, this observation is a motivating factor for the conjecture of Nielsen (mentioned in Section~\ref{sec:intro}) regarding conditions for catalytic majorization.

Our functions $f_r$ are known to be additive Schur-convex
functions.  This can be verified with the aid of the following fact (from, e.g., \cite{Mar79}):  
A differentiable function $f(x)$ is Schur-convex if and only
if it is invariant to permutations of the components of $x$ and
\[ (x_i-x_j)\left( \frac{\partial f}{\partial x_i}
    - \frac{\partial f}{\partial x_j} \right) \geq 0 \]
for any pair of indices $i,j$.  (For our application, the definition of additive Schur-convexity must be modified slightly to allow for the fact that $f_r$ can be $\infty$ when $r \leq 0$, but this does not cause much difficulty.)

Clearly we have that $f_r(x)$ is continuous in $r$ when $r \neq 0,1$.  However we are really more interested in the inequality $f_r(x) < f_r(y)$ than the actual values of the functions.  It turns out that this inequality is continuous in $r$ in a certain sense for all $r \in \myreal$.  Specifically, if for any neighborhood of $r$ there exists an $r'$ in the neighborhood such that $f_{r'}(x) \leq f_{r'}(y)$, then $f_r(x) \leq f_r(y)$.

This continuity of the $f_r$ inequalities in $r$ can be exhibited as follows.  Let $\tilde{f}_r(x) = (1/r(r-1)) \ln \sum_{i=1}^d x_i^r$, for $r \neq 0,1$.  Clearly $\tilde{f}_r(x) < \tilde{f}_r(y)$ is equivalent to $f_r(x) < f_r(y)$ for $r \neq 0,1$.  For the case $r=1$, we note that the limit of $\tilde{f}_r(x)$ as
$r$ approaches $1$ is equal to $f_1(x)$.  On the other hand, in the limit as $r \rightarrow 0$, $\tilde{f}_r(x)$ goes to $\pm \infty$ (depending on which direction the limit is from).  However, the difference $\tilde{f}_r(y)- \tilde{f}_r(x)$ converges to $(f_0(y)-f_0(x))/d$ if neither $x$ nor $y$ contains components equal to zero.  Thus, if we use the limiting values when $r \neq 0,1$ then the difference $\tilde{f}_r(y)- \tilde{f}_r(x)$ is continuous in $r$ over all of $\myreal$ when neither $x$ nor $y$ contains components equal to zero.  Given this fact and our main theorem, it may be convenient in some cases to assess whether $x \prec_T y$ by examining a plot of $\tilde{f}_r(y)- \tilde{f}_r(x)$ as a function of $r$.

There are other additive Schur-convex functions that are not members of our $f_r$ family of functions; examples are given in, e.g., \cite{Nie99a}.  We note in particular the functions
\[ x \mapsto \ln x_1^{\downarrow}, \]
\[ x \mapsto -\ln x_d^{\downarrow}, \]
and
\[ x \mapsto -\ln \left|\supp x \right|, \]
where $\left|\supp x \right|$ is the number of nonzero components of $x$.  From our main theorem and the above discussion, it must be the case that if for some $x$ and $y$ we have $f_r(x) \leq f_r(y)$ for all $r \in \myreal$, then $f(x) \leq f(y)$ for any additive Schur-convex function $f$.  This can be easily verified directly for the examples above.


\section{Order-Free Characterization of Majorization}  \label{sec:order-free}
The usual definition of majorization involves arranging the components of the vectors in decreasing order.  This definition appears to be inconvenient to work with when considering catalytic majorization.  Thus we rely on the following well-known equivalence (see, e.g., \cite{Bha97}), where the notation $(c)^{+}$ means the positive portion of $c$; that is, $(c)^{+} =\max(c,0)$.
\begin{proposition}
For $d$-dimensional vectors $x$ and $y$ the following are equivalent:
\begin{itemize}
\item[(i)] $x \prec y$;
\item[(ii)] $\sum_{i=1}^d x_i = \sum_{i=1}^d y_i$ and for all $t \in \myreal$,
  \[ \sum_{i=1}^d (x_i-t)^{+} \leq \sum_{i=1}^d (y_i-t)^{+}; \]
\item[(iii)] $\sum_{i=1}^d x_i = \sum_{i=1}^d y_i$ and for all $t \in \myreal$,
  \[ \sum_{i=1}^d (t-x_i)^{+} \leq \sum_{i=1}^d (t-y_i)^{+}. \]
\end{itemize}
\end{proposition}

In the sufficiency portion of the proof of our main result, we use the following formulation.  If $x$ and $y$ are $d$-dimensional probability vectors and $z$ is an $\ell$-dimensional vector, then $x \otimes z \prec y \otimes z$ if and only if for all $t \in \myreal$ we have
\begin{equation}  \label{eq:order_free_cm}
  \sum_{i=1}^d \sum_{j=1}^{\ell} (x_i z_j-t)^{+}
  \leq \sum_{i=1}^d \sum_{j=1}^{\ell} (y_i z_j-t)^{+}.
\end{equation}

\section{Necessity}
We first establish the necessity direction of our main theorem: under the hypothesis of Theorem~\ref{thm:main}, $x \prec_T y$ implies $f_r(x) < f_r(y)$ for all $r \in \myreal$.  This almost follows from the discussion of Schur-convex functions in Section~\ref{sec:asc}.  However, to show that the inequalities must all be strict, we use the following result of Daftuar and the author \cite{Daf01}:
\begin{corollary}  \label{cor:key}
Let $x =(x_1,\ldots, x_d)$ and $y=(y_1,\ldots, y_d)$ be $d$-dimensional probability vectors.  Let $T(y)$ denote the set of all $d$-dimensional probability vectors $x'$ such that $x' \prec_T y$.  Suppose that $x \prec_T y$ and $y_1^{\downarrow} > x_1^{\downarrow}$ and $y_d^{\downarrow} < x_d^{\downarrow}$.  Then $x$ is in the interior of $T(y)$ (relative to the space of probability vectors).
\end{corollary}

\emph{Sketch of proof of necessity portion of Theorem~\ref{thm:main}:}
Suppose $x \prec_T y$.  Because each $f_r$ is additive Schur-convex, we must have $f_r(x) \leq f_r(y)$ for all $r \in \myreal$.  However, we must show that equality cannot hold.  We assume that the components of both $x$ and $y$ are in decreasing order.  Without loss of generality, we may assume that no component of $x$ is equal to any component of $y$, as removing such a component from both $x$ and $y$ does not change the status of the catalytic majorization relation between them, nor does it affect the $f_r$ inequalities.  Using this assumption and the fact that $f_r(x) \leq f_r(y)$ as $r$ approaches $+\infty$ and $-\infty$, we find that we may assume $x_1 < y_1$ and $x_d > y_d$.  Let $w$ be the $d$-dimensional vector $(1,0,\ldots,0,-1)$.
From Corollary~\ref{cor:key}, $x$ is not on the boundary of $T(y)$, so for a sufficiently small $\epsilon>0$, we have $x+\epsilon w \prec_T y$.
It is straightforward to verify that $f_r(x) < f_r(x+\epsilon w)$ with this $\epsilon$ for all $r \in \myreal$.  But $f_r(x+\epsilon w) \leq f_r(y)$ since $x+\epsilon w \prec_T y$, so necessity is established.
\myqed

\section{Sufficiency Outline}  \label{sec:sufficiency_outline}
We now outline how the sufficiency portion of Theorem~\ref{thm:main} can be proved.  That is, under the hypothesis of Theorem~\ref{thm:main}, we outline how to show that if $f_r(x) < f_r(y)$ for all $r \in \myreal$ then there exists a catalyst vector $z$ such that $x \otimes z \prec y \otimes z$.

For convenience, we assume that the components of $x$ and $y$ are arranged in decreasing order.  As in the necessity proof, we may assume without loss of generality that $x_1 < y_1$ and $x_d > y_d$.  We allow catalyst to vectors to be unnormalized.


We argue that we need only consider vectors $x$ and $y$ whose components are all nonzero.  The constraint $x_d > y_d$ implies that the components of $x$ are all nonzero.  For $y$, we use the following claim.
\begin{claimnn}
If $x$ and $y$ satisfy the $f_r$ inequalities, and $x_1 < y_1$ and $x_d > y_d$, then there exists a $y'$ (with components assumed to be in decreasing order) whose components are all nonzero, with $x_1 < y'_1$ and $x_d > y'_d$, such that $y' \prec y$, and the $f_r$ inequalities are satisfied with $y$ replaced by $y'$.
\end{claimnn}
This claim is straightforward (but slightly tedious) to verify; we provide a proof in the Appendix.  Suppose the sufficiency portion of Theorem~\ref{thm:main} is shown to hold for probability vectors whose components are all nonzero.  Then given $x$ and $y$ satisfying the $f_r$ inequalities, we can find $y'$ with the properties stated in the claim and conclude that there exists a $z$ such that $x \otimes z \prec y' \otimes z$.  But $y' \prec y$ implies $y' \otimes z \prec y \otimes z$, so we have $x \otimes z \prec y \otimes z$ as desired.  Thus we henceforth assume that all components of $y$ (and of $x$) are nonzero.


The sufficiency proof can be divided into steps as follows.  In these steps, we assume $x$ and $y$ are $d$-dimensional probability vectors with all nonzero components, the components of $x$ and $y$ are arranged in decreasing order, and $x_1 < y_1$ and $x_d > y_d$.

\textbf{Step 1.} Show that if the $f_r$ inequalities hold for all $r \geq 0$, then there exists a continuous, decreasing function
$z_{+}: [0, \infty) \rightarrow (0,1]$ with $z_{+}(0)=1$, along with a constant $s_{+}$, such that
  \begin{enumerate}
  \item[(i)] for all $t \in (0,y_1)$,
  \begin{equation}  \label{eq:1ci}
    \sum_{i=1}^d \int_0^{\infty} (x_i z_{+}(s)-t)^{+}\,ds
      < \sum_{i=1}^d \int_0^{\infty} (y_i z_{+}(s)-t)^{+}\,ds;
  \end{equation}
  \item[(ii)] if $s \geq s_{+}$ then $z_{+}(s) = z_{+}(s_{+}) e^{-(s-s_{+})}$; and
  \item[(iii)] the function $z_{+}$ satisfies a Lipschitz condition; that is, there exists a $K$ such that if $s_1,s_2 \geq 0$ then $|z_{+}(s_1)-z_{+}(s_2)| \leq K|s_1-s_2|$.
  \end{enumerate}

\textbf{Step 2.} Show that if the $f_r$ inequalities hold for all $r \leq 0$, then there exists a continuous, increasing function $z_{-}: [0, \infty) \rightarrow [1,\infty)$ with $z_{-}(0)=1$, along with a constant $s_{-}$, such that 
  \begin{enumerate}
  \item[(i)] for all $t \in (y_d,\infty)$,
  \[ \sum_{i=1}^d \int_0^{\infty} (t-x_i z_{-}(s))^{+}\,ds
      < \sum_{i=1}^d \int_0^{\infty} (t-y_i z_{-}(s))^{+}\,ds; \]
  \item[(ii)] if $s \geq s_{-}$ then $z_{-}(s) = z_{-}(s_{-}) e^{s-s_{-}}$; and
  \item[(iii)] the function $z_{-}$ satisfies a Lipschitz condition on the interval $[0, s_{-}]$.
  \end{enumerate}

\textbf{Step 3.} Show that if $z_{+}$ and $z_{-}$ satisfy the conditions given in Steps 1 and 2, then there exists a continuous, decreasing, positive function $z_{*}(s)$ on an interval $[0,a]$, with $z_{*}(0)=1$, such that
  \begin{enumerate}
  \item[(i)] for all $t \in (y_d z_{*}(a),y_1)$,
   \begin{equation}  \label{eq:step3}
    \sum_{i=1}^d \int_0^a (x_i z_{*}(s)-t)^{+}\,ds
      < \sum_{i=1}^d \int_0^a (y_i z_{*}(s)-t)^{+}\,ds;
   \end{equation}
  and
  \item[(ii)] the function $z_{*}$ satisfies a Lipschitz condition on $[0,a]$.
  \end{enumerate}

\textbf{Step 4.} Show that if $z_{*}$ satisfies the conditions given in Step~3, then there exists a finite-dimensional $z$ for which
$x \otimes z \prec y \otimes z$.

Clearly, the results to be proved in Steps 1 through 4 together imply the sufficiency portion of Theorem~\ref{thm:main}.

\section{Overview of Steps 1 and 2}
Here we present an outline of how Steps 1 and 2 can be completed.  The presentation here is somewhat less detailed than that of the other portions of the proof, but we hope that this outline provides a reasonable indication of our general strategy and of the methods involved.

First we provide some motivation for why the results in these steps are useful.  The basic idea is that conditions~(i) of Steps 1 and 2 are reminiscent of the order-free conditions for majorization discussed in Section~\ref{sec:order-free}.  In fact one could regard condition~(i) of Step~1 as implying that $x$ is catalytically majorized by $y$ with catalyst $z_{+}$, in an appropriately generalized sense.  However, we note that this condition by itself does not imply that $x \prec_T y$.  (A similar observation is made in \cite{Daf04,Aub07}.)  Conditions~(ii) of Steps 1 and 2 together provide a way to combine the results of Steps 1 and 2, as will be clear in carrying out Step~3.  We provide further thoughts related to conditions~(i) of Steps 1 and 2 in Section~\ref{sec:discussion}.

In Step~1, we first find a function $\tilde{z}_{+}$ that satisfies condition~(i) without regard for condition~(ii).  A key idea is to consider functions of the form
\[ \tilde{z}_{+}(s) = e^{-s^{1/n}}, \]
where $n$ is a positive integer.  It turns out that if for a given $x$ and $y$ the $f_r$ inequalities are satisfied for $r \geq 0$, then for sufficiently large $n$ condition~(i) holds with this $\tilde{z}_{+}$ substituted for $z_{+}$.

To establish this result, we parameterize $t$ as $t = e^{-n/r}$; the range $r \in (0,\infty)$ corresponds to the range $t \in (0,1)$.  For a given $r$ there is a correspondence between the inequality (\ref{eq:1ci}) for $t = e^{-n/r}$ and the $f_r$ inequality.  Specifically, the following holds:  If for some $r>0$ the $f_r$ inequality is satisfied, then for all sufficiently large $n$ the inequality (\ref{eq:1ci}) holds for $t = e^{-n/r}$ and $\tilde{z}_{+}(s) = e^{-s^{1/n}}$.  In other words, given $r>0$, if $f_r(x) < f_r(y)$ then for all sufficiently large $n$ we have
\begin{equation}  \label{eq:cond1a}
 \sum_{i=1}^d \int_0^{\infty} (x_i e^{-s^{1/n}} - e^{-n/r})^{+}\,ds
    < \sum_{i=1}^d \int_0^{\infty} (y_i e^{-s^{1/n}} - e^{-n/r})^{+}\,ds.
\end{equation}
We require the existence of an $n$ such that (\ref{eq:cond1a}) holds for all $r>0$, so it is not enough just to show that the above implication holds for each $r>0$.

Our strategy in showing such an $n$ exists involves dividing the values of $r$ into five regions:  regions 1, 3, and 5 corresponds conceptually to neighborhoods of $0$, $1$, and $\infty$, respectively, while regions 2 and 4 fill in the gaps.  More precisely, the steps in the strategy are as follows.

\textbf{Region~5.}
Show that the inequality $y_1 > x_1$ implies that there is an interval $R_5 = [r_4,\infty)$ for some $r_4 \in (1,\infty)$, along with a positive integer $n_5$, such that if $r \in R_5$ and $n \geq n_5$, then (\ref{eq:cond1a}) holds.

\textbf{Region~3.}
Show that the $f_r$ inequality with $r=1$ implies that there is an interval $R_3 = [r_2,r_3]$, where $r_2 \in (0,1)$ and $r_3 \in (1,r_4)$, along with a positive integer $n_3$, such that if $r \in R_3$ and $n \geq n_3$, then (\ref{eq:cond1a}) holds.

\textbf{Region~1.}
Show that the $f_r$ inequality with $r=0$ implies that there is an interval $R_1 = (0,r_1]$ for some $r_1 \in (0,r_2)$, such that if $r \in R_1$ and $n \geq 1$, then (\ref{eq:cond1a}) holds.  Note the condition $n \geq 1$; unlike the other regions, $n$ does not need to be large here.

\textbf{Region~2.}
The interval $R_2$ is $[r_1,r_2]$.  The task for this step is to show that the $f_r$ inequalities for $r \in R_2$ imply that there exists a positive integer $n_2$ such that if $r \in R_2$ and $n \geq n_2$, then (\ref{eq:cond1a}) holds.

\textbf{Region~4.}
Similarly, the interval $R_4$ is $[r_3,r_4]$.  The task for this step is to show that the $f_r$ inequalities for $r \in R_4$ imply that there exists a positive integer $n_4$ such that if $r \in R_4$ and $n \geq n_4$, then (\ref{eq:cond1a}) holds.

We proceed with some mathematical preliminaries that we found useful for completing the region steps above.  For convenience we let
\[ \alpha(c,t,n)= \int_0^{\infty} (ce^{-s^{1/n}} - t)^{+} \,ds, \]
where $c \in (0,1)$ and $n \geq 1$ and $t>0$.

We first observe that for $n \geq 1$ and $s \geq 0$,
\[ \int e^{-s^{1/n}}\,ds
    = -n!\,e^{-s^{1/n}} \sum_{j=0}^{n-1} \frac{1}{j!} s^{j/n}. \]
This can be verified by taking the derivative of the right side with respect to $s$.

Suppose $c \geq t$.  Then
\begin{align}
  \alpha(c,t,n)
  & = \int_0^{(\ln \frac{c}{t})^n} (ce^{-s^{1/n}} - t) \,ds \nonumber \\
  & = -cn! \left[ e^{-s^{1/n}} \sum_{j=0}^{n-1} \frac{1}{j!} s^{j/n}
        \right]_0^{(\ln \frac{c}{t})^n} - \left( \ln \frac{c}{t} \right)^n t \nonumber \\
  & = -cn! \left(\frac{t}{c} \sum_{j=0}^{n-1} \frac{1}{j!}
        \left( \ln \frac{c}{t} \right)^j - 1 \right)
        - \left( \ln \frac{c}{t} \right)^n t \nonumber \\
  & = cn! - tn! \left(\sum_{j=0}^{n-1} \frac{1}{j!}
        \left( \ln \frac{c}{t} \right)^j
        + \frac{1}{n!} \left( \ln \frac{c}{t} \right)^n \right) \nonumber \\
  & = cn! - tn! \sum_{j=0}^n \frac{1}{j!}
        \left( \ln \frac{c}{t} \right)^j. \label{eq:alpha_expr1}
\end{align}
Noting that $\sum_{j=0}^{\infty} \frac{1}{j!} (\ln \frac{c}{t})^j = c/t$, it follows from (\ref{eq:alpha_expr1}) that
\[
  \alpha(c,t,n) = tn! \sum_{j=n+1}^{\infty} \frac{1}{j!}
    \left( \ln \frac{c}{t} \right)^j.
\]
Substituting $t=e^{-n/r}$ gives
\[
  \alpha(c,e^{-n/r},n)
   = e^{-n/r} n! \sum_{j=n+1}^{\infty} \frac{1}{j!}
    \left( \ln c + \frac{n}{r} \right)^j.
\]
We found this form of $\alpha$ to be a useful starting point for completing the region~5 step.

Returning to (\ref{eq:alpha_expr1}), we substitute $t=e^{-n/r}$ and find that
\begin{align}
  \alpha(c,e^{-n/r},n)
  & = cn! - e^{-n/r} n! \sum_{j=0}^n \frac{1}{j!}
    \left( \ln c + \frac{n}{r} \right)^j \nonumber \\
  & = cn! - e^{-n/r} n! \sum_{j=0}^n \frac{1}{j!}
    \sum_{k=0}^j {j \choose k} (\ln c)^k \left(\frac{n}{r} \right)^{j-k} \nonumber \\
  & = cn! - e^{-n/r} n! \sum_{k=0}^n (\ln c)^k
    \sum_{j=k}^n \frac{1}{j!} {j \choose k} \left(\frac{n}{r} \right)^{j-k} \nonumber \\
  & = cn! - e^{-n/r} n! \sum_{k=0}^n \frac{1}{k!} (\ln c)^k
    \sum_{j=k}^n \frac{1}{(j-k)!} \left(\frac{n}{r} \right)^{j-k} \nonumber \\
  & = cn! - e^{-n/r} n! \sum_{k=0}^n \frac{1}{k!} (\ln c)^k
    \sum_{j=0}^{n-k} \frac{1}{j!} \left(\frac{n}{r} \right)^j, \label{eq:alpha_region12}
\end{align}
where in the last step we have replaced $j-k$ with $j$.  We found the form of $\alpha$ given by (\ref{eq:alpha_region12}) to be a useful starting point for completing the region~1 and region~2 steps.

Now note that $\sum_{k=0}^{\infty} \frac{1}{k!} (\ln c)^k \sum_{j=0}^{\infty} \frac{1}{j!} (\frac{n}{r})^j = ce^{n/r}$.  Thus we can conclude from (\ref{eq:alpha_region12}) that
\begin{equation}  \label{eq:alpha_region34}
 \alpha(c,e^{-n/r},n) = e^{-n/r} n! \sum_{k=0}^{\infty} \frac{1}{k!} (\ln c)^k
    \sum_{j=\max(n+1-k,0)}^{\infty} \frac{1}{j!} \left(\frac{n}{r} \right)^j.
\end{equation}
We found this form of $\alpha$ to be a useful starting point for completing the region~3 and region~4 steps.

As an example we discuss our approach for the region~4 step in some detail.  We start with (\ref{eq:alpha_region34}) where we replace $j$ with $j+n+1-k$:
\begin{align}
  \alpha(c,e^{-n/r},n)
  & = e^{-n/r} n! \sum_{k=0}^{\infty} \frac{1}{k!} (\ln c)^k
      \sum_{j=\max(0,k-n-1)}^{\infty} \frac{1}{(n+1+j-k)!}
      \left(\frac{n}{r} \right)^{n+1+j-k} \nonumber \\
  & = e^{-n/r} n! \sum_{k=0}^{\infty} \frac{1}{k!} r^k (\ln c)^k
      \sum_{j=\max(0,k-n-1)}^{\infty} \frac{n^{n+1+j-k}}{(n+1+j-k)!}
      r^{-n-1-j} \nonumber \\
  & = e^{-n/r} n! \frac{n^{n+1}}{r^{n+1}(n+1)!} \sum_{k=0}^{\infty}
      \frac{1}{k!} (\ln c^r)^k \sum_{j=\max(0,k-n-1)}^{\infty}
      \frac{(n+1)!\,n^{j-k}}{(n+1+j-k)!} r^{-j}  \label{eq:alpha_region4}
\end{align}
%
We present a sequence of successively refined results regarding this expression.
It is not difficult to show that for fixed $j$ and $k$,
\[ \lim_{n \rightarrow \infty} \frac{(n+1)!\,n^{j-k}}{(n+1+j-k)!} = 1. \]
By crudely bounding the difference between $\frac{(n+1)!\,n^{j-k}}{(n+1+j-k)!}$ and $1$ for large $n$, it can be shown that for fixed $r>1$ and fixed $k$,
\[ \lim_{n \rightarrow \infty} \sum_{j=\max(0,k-n-1)}^{\infty}
   \frac{(n+1)!\,n^{j-k}}{(n+1+j-k)!}r^{-j} = \sum_{j=0}^{\infty} r^{-j}
   = \frac{r}{r-1}. \]
By instead carefully bounding the quantities involved in the double sum in (\ref{eq:alpha_region4}) it can be shown that for a fixed $c \in (0,1)$ and a fixed $r>1$,
\[ \lim_{n \rightarrow \infty} \sum_{k=0}^{\infty}
      \frac{1}{k!} (\ln c^r)^k \sum_{j=\max(0,k-n-1)}^{\infty}
      \frac{(n+1)!\,n^{j-k}}{(n+1+j-k)!} r^{-j}
  = \frac{r}{r-1} \sum_{k=0}^{\infty} \frac{1}{k!} (\ln c^r)^k
  = \frac{r}{r-1} c^r. \]
Continuing along these lines, it can be shown that for a fixed $r>1$, if $\sum_{i=1}^d x_i^r < \sum_{i=1}^d y_i^r$ then for $n$ sufficiently large (\ref{eq:cond1a}) holds.  Finally, given $r_3$ and $r_4$ with $1<r_3<r_4$, it can be shown that if the $f_r$ inequality holds for all $r \in [r_3,r_4]$, then there exists an $n_4$ such that if $n \geq n_4$ then (\ref{eq:cond1a}) holds for all $r \in [r_3,r_4]$.  Note that only this last result is needed to complete the region~4 step; the sequence of results is given to provides some intuition as to why the last result holds, as well as suggesting how to establish it.

After all of the region steps are completed, it follows quickly that condition~(i) of Step~1 is satisfied by $\tilde{z}_{+}(s) = e^{-s^{1/n}}$ for sufficiently large $n$.  We now explain how a $z_{+}$ can be constructed that also satisfies condition~(ii).  We start with a straightforward result that shows that whether (\ref{eq:1ci}) holds for a given $t$ (or range of $t$) depends only on a portion of the function $z_{+}$.

\begin{claimnn}
In the context of Step~1, let $z_{+}^{(0)}$ and $z_{+}^{(1)}$ be continuous, positive, decreasing functions on $[0,\infty)$ with $z_{+}^{(0)}(0) = z_{+}^{(1)}(0) = 1$.
\begin{itemize}
\item[(i)] Suppose $z_{+}^{(0)}(s) = z_{+}^{(1)}(s)$ for all $s \in [0,s_0]$ for some $s_0$, and that (\ref{eq:1ci}) holds for $z_{+}^{(0)}$ when $t \in [y_1 z_{+}^{(0)}(s_0), y_1)$.  Then (\ref{eq:1ci}) holds for $z_{+}^{(1)}$ when $t$ is in this same interval.
\item[(ii)] Fix $t$ and let $[s_1,s_2]$ be the interval for which $z_{+}^{(0)}(s)$ is in $[t/y_1, t/y_d]$.  Suppose for a given $\Delta$ we have $z_{+}^{(1)}(s+\Delta) = z_{+}^{(0)}(s)$ for all $s \in [s_1,s_2]$ (where necessarily $s_1+\Delta \geq 0$), and that (\ref{eq:1ci}) holds for $z_{+}^{(0)}$ with our specific $t$.  Then (\ref{eq:1ci}) holds for $z_{+}^{(1)}$ and $t$.
\end{itemize}
\end{claimnn}

\emph{Sketch of proof.}  We outline a proof of part~(ii); part~(i) can be proved by similar means.  Observe that if $c$ is a component of $x$ or $y$, then $c z_{+}^{(1)}(s) \geq t$ when $s \leq s_1+\Delta$, and $c z_{+}^{(1)}(s) \leq t$ when $s \geq s_2+\Delta$.  We therefore have
\begin{align}
  \sum_{i=1}^d \int_0^{\infty} (x_i z_{+}^{(1)}&(s) - t)^{+} \,ds \nonumber \\
  & = \sum_{i=1}^d \left(
        \int_0^{s_1+\Delta} (x_i z_{+}^{(1)}(s) - t) \,ds
        + \int_{s_1+\Delta}^{s_2+\Delta} (x_i z_{+}^{(1)}(s) - t)^{+} \,ds
        \right) \nonumber \\
  & = \sum_{i=1}^d \left(-(s_1+\Delta)t + x_i \int_0^{s_1+\Delta} z_{+}^{(1)}(s)\,ds
        + \int_{s_1}^{s_2} (x_i z_{+}^{(0)}(s) - t)^{+} \,ds
        \right) \nonumber \\
  & = -(s_1+\Delta)td + \int_0^{s_1+\Delta} z_{+}^{(1)}(s)\,ds 
        + \sum_{i=1}^d \int_{s_1}^{s_2} (x_i z_{+}^{(0)}(s) - t)^{+} \,ds
        \label{eq:intsplit1}
\end{align}
and similarly
\begin{equation}  \label{eq:intsplit2}
 \sum_{i=1}^d \int_0^{\infty} (x_i z_{+}^{(0)}(s) - t)^{+} \,ds
   = \int_0^{s_1} z_{+}^{(0)}(s) \,ds - s_1 td
     + \sum_{i=1}^d \int_{s_1}^{s_2} (x_i z_{+}^{(0)}(s) - t)^{+} \,ds.
\end{equation}
We see from (\ref{eq:intsplit1}) and (\ref{eq:intsplit2}) that the difference
\[ \sum_{i=1}^d \int_0^{\infty} (x_i z_{+}^{(1)}(s) - t)^{+} \,ds 
   - \sum_{i=1}^d \int_0^{\infty} (x_i z_{+}^{(0)}(s) - t)^{+} \,ds \]
does not depend on $x$.  The analogous expression with $x$ replaced by $y$ has the same value.  Thus (with our specific $t$), condition (\ref{eq:1ci}) holds for $z_{+}^{(0)}$ if and only if (\ref{eq:1ci}) holds for $z_{+}^{(1)}$.
\myqed

To produce the desired $z_{+}$, we start with $\tilde{z}_{+}(s) = e^{-s^{1/n_0}}$, where $n_0$ is chosen to be large enough that $\tilde{z}_{+}$ satisfies condition (i) of Step~1.  Let $z_{+}(s) = \tilde{z}_{+}(s)$ when $s \in [0,s_0]$, where $s_0$ is chosen to be large enough that (\ref{eq:1ci}) holds for all $t \in [e^{-n_0/r_1},y_1)$ regardless of how we specify the rest of $z_{+}$; the existence of such an $s_0$ is implied by part (i) of the preceding claim.  The remaining $t$ fall in the interval $(0, e^{-n_0/r_1})$ which, with $t$ parameterized as $t=e^{-n_0/r}$, corresponds to $r \in R_1$.  Recall that in this region it was not necessary for $n$ to be large for (\ref{eq:1ci}) to be satisfied by $\tilde{z}_{+}(s) = e^{-s^{1/n}}$.

The idea is to extend $z_{+}$ beyond $s_0$ in such a way that $z_{+}$ gradually changes from the form $e^{-s^{1/n_0}}$ to the form $e^{-s}$.  We outline how this can be accomplished.  We extend $z_{+}$ with sections from functions of the form $e^{-s^{1/n}}$ (with offsets chosen to make $z_{+}$ continuous), where $n$ decreases from section to section and is now no longer necessarily an integer.  We make each such section long enough that the value of $z_{+}$ decreases by at least a factor of $y_1/y_d$ within the section.  Part~(ii) of our claim then implies that to verify (\ref{eq:1ci}) for a given $t$ we need only consider the value of $z_{+}$ in at most two sections.  The starting point for doing this is to show that (\ref{eq:cond1a}) holds for noninteger $n \geq 1$ for $t$ in region~1.  Then it must be checked that when the decrement to $n$ is small enough, (\ref{eq:1ci}) continues to hold for the relevant values of $t$.  Finally, it must be shown that for sufficiently many decrements we can reach a segment with $n=1$.  The segment with $n=1$ is used for the remaining $s$, and part~(ii) of our claim is again used to show that (\ref{eq:1ci}) holds.  At this point $z_{+}$ can be verified to satisfy all parts required by Step~1.

Step~2 is largely similar to Step~1.  A key idea in Step~2 is to consider functions of the form $\tilde{z}_{-}(s) = e^{s^{1/n}}$ where $n$ is a positive integer.  If the $f_r$ inequalities are satisfied for all $r \leq 0$, then for sufficiently large $n$ condition (i) in Step~2 holds with this $\tilde{z}_{-}$.  As in Step~1, our strategy for showing this involves five regions of values of $r$, in this case with regions $1'$, $3'$, and $5'$ corresponding conceptually to neighborhoods of $0$, $-1$, and $-\infty$.  A $z_{-}$ satisfying condition (ii) of Step~2 is constructed analogously to the construction of the $z_{+}$ satisfying condition (ii) of Step~1.

%
%
\ignore{
The significance of this inequality is that it is reminiscent of a well-known equivalent condition for majorization: $x \prec y$ if and only if $\sum x_i =\sum y_i$ and $\sum (x_i-t)^{+} \leq \sum (y_i-t)^{+}$ for all $t \in R$.  It turns out that we can use the function $z_{+}(s) = e^{-s^{1/n}}$ if $n$ is large enough.  In fact, only the $f_t$ inequalities for $t \in [0,\infty)$ are needed to complete this step.  This suggests that if we allow $z$ to be an infinite sequence, we could have $x \otimes z \prec y \otimes z$ (in a generalized sense) without the $f_t$ inequalities for negative $t$.  This is true, but it turns out that such a $z$ sequence does not imply the existence of a finite dimensional $z$ satisfying $x \otimes z \prec y \otimes z$.
}

\section{Step 3}
The function $z_{*}$ in this step may be regarded as a continuous precursor to a finite-dimensional catalyst vector $z$.  The construction of $z_{*}$ is roughly as follows.  The function $z_{+}$ from Step 1 is used for the left side of the interval on which $z_{*}$ is defined, and the function $z_{-}$ from Step 2 is reflected horizontally, scaled appropriately, and used for the right side of the interval on which $z_{*}$ is defined.  There is a region in the center, in which $z_{*}$ is equal to a decaying exponential function, where the two halves coincide.  The center region allows the transition from the submajorization-like condition involving $z_{+}$ to the supermajorization-like condition involving $z_{-}$ to be made.


We proceed with the details.  Let $a=s_{+}+s_{-}+ \ln(y_1/y_d)$.  The function $z_{*}$ is defined on the interval $[0,a]$.  We let
\begin{equation}  \label{eq:zstar_defn}
  z_{*}(s) =
  \begin{cases}
    z_{+}(s) & \text{if $s \in [0, a-s_{-}]$;} \\
    \frac{y_d}{y_1} \frac{z_{+}(s_{+})}{z_{-}(s_{-})} z_{-}(a-s)
       & \text{if $s \in (a-s_{-},a]$.}
  \end{cases}
\end{equation}

Observe that if $s \in [s_{+}, a-s_{-}]$ then
\begin{align*}
  \frac{y_d}{y_1} \frac{z_{+}(s_{+})}{z_{-}(s_{-})} z_{-}(a-s)
  & = \frac{y_d}{y_1} \frac{z_{+}(s_{+})}{z_{-}(s_{-})}
        z_{-}(s_{-}) e^{a-s_{-}-s} \\
  & = \frac{y_d}{y_1} z_{+}(s_{+}) e^{s_{+} + \ln(y_1/y_d) - s} \\
  & = z_{+}(s_{+}) e^{-(s-s_{+})} \\
  & = z_{+}(s),
\end{align*}
and therefore when $s \in [s_{+},a-s_{-}]$, either part of (\ref{eq:zstar_defn}) can be used.  This is the aforementioned region where the two parts of $z_{*}(s)$ coincide.  Also observe that if $s \in [s_{+},a]$ then $z_{*}(s) = z_{*}(a)z_{-}(a-s)$.  Note that $z_{*}$ is continuous and decreasing and satisfies a Lipschitz condition.

Let $t^{*} = y_1 z_{*}(a-s_{-})$.  Observe that also
\[ t^{*} = y_1 z_{*}(a-s_{-}) = y_1 z_{+}(s_{+}) e^{-(a-s_{-}-s_{+})}
   = y_1 z_{+}(s_{+}) e^{- \ln(y_1/y_d)} = y_d z_{+}(s_{+}). \]

Suppose $c$ is a component of $x$ or $y$, so that in particular $0 < c \leq y_1$.  If $s \in [a-s_{-},a]$ and $t \geq t^{*}$ then
\[ c z_{*}(s)-t \leq y_1 z_{*}(a-s_{-}) - t^{*} = 0, \]
and similarly if $s \geq a-s_{-}$ and $t \geq t^{*}$ then $c z_{+}(s)-t \leq 0$.  Since $z_{*}(s)=z_{+}(s)$ when $s \in [0,a-s_{-}]$, for all $t \geq t^{*}$ we have
\[ \int_0^a (cz_{*}(s)-t)^{+} \,ds = \int_0^{\infty} (cz_{+}(s)-t)^{+} \,ds. \]
Now the result from Step 1 implies that if $t \in [t^{*},y_1)$ then
\begin{equation} \label{eq:cmaj1}
 \sum_{i=1}^d \int_0^a (x_i z_{*}(s)-t)^{+} \,ds
   < \sum_{i=1}^d \int_0^a (y_i z_{*}(s)-t)^{+} \,ds.
\end{equation}

Next we find an analogous result that uses the result from Step~2.  The derivation is slightly more involved due to the fact that $z_{-}$ is reflected and scaled to form the right half of $z_{*}$.  Suppose again that $c$ is a component of $x$ or $y$, so that in particular $c \geq y_d$.  If $s \in [0,s_{+}]$ and $t \leq t^{*}$ then
\[ t-cz_{*}(s) \leq t^{*}-y_d z_{*}(s_{+}) = 0. \]
Thus we have
\begin{align*}
  \int_0^a (t-cz_{*}(s))^{+} \,ds
  & = \int_{s_{+}}^a (t-cz_{*}(s))^{+} \,ds \\
  & = \int_{s_{+}}^a (t-cz_{*}(a)z_{-}(a-s))^{+} \,ds \\
  & = z_{*}(a) \int_{s_{+}}^a \left(\frac{t}{z_{*}(a)}
         - cz_{-}(a-s) \right)^{+} ds \\
  & = z_{*}(a) \int_0^{a-s_{+}} \left(\frac{t}{z_{*}(a)}
         - cz_{-}(u) \right)^{+} du,
\end{align*}
where we have made the substitution $u=a-s$.

For $u \geq a-s_{+}$ and $t \leq t^{*}$ we have
\[ \frac{t}{z_{*}(a)} - c z_{-}(u)
  \leq \frac{t^{*}}{z_{*}(a)} - y_d z_{-}(a-s_{+})
  = \frac{y_d z_{+}(s_{+})}{z_{*}(a)} - y_d \frac{z_{*}(s_{+})}{z_{*}(a)}
  = 0. \]
Therefore
\[ z_{*}(a) \int_0^{a-s_{+}} \left(\frac{t}{z_{*}(a)}
         - cz_{-}(u) \right)^{+} du
  = z_{*}(a) \int_0^{\infty} \left(\frac{t}{z_{*}(a)}
         - cz_{-}(u) \right)^{+} du  \]
and so
\[ \int_0^a (t-cz_{*}(s))^{+} \,ds
   = z_{*}(a) \int_0^{\infty} \left(\frac{t}{z_{*}(a)}
         - cz_{-}(u) \right)^{+} du. \]
Thus the result from Step 2 implies that if $t \in (y_d z_{*}(a), t^{*}]$ then
\begin{equation} \label{eq:cmaj2}
  \sum_{i=1}^d \int_0^a (t-x_i z_{*}(s))^{+} \,ds
   < \sum_{i=1}^d \int_0^a (t-y_i z_{*}(s))^{+} \,ds.
\end{equation}

Finally we show that for any given $t$, conditions (\ref{eq:cmaj1})
and (\ref{eq:cmaj2}) are equivalent.  Observe that $c = (c)^{+} - (-c)^{+}$ for any real $c$; thus
\begin{align*}
 \sum_{i=1}^d \int_0^a (x_i z_{*}(s)-t)^{+} \,ds
 & = \sum_{i=1}^d \int_0^a (x_i z_{*}(s)-t + (t-x_i z_{*}(s))^{+}) \,ds \\
 & = \int_0^a z_{*}(s)\,ds - tad
       + \sum_{i=1}^d \int_0^a (t-x_i z_{*}(s))^{+} \,ds.
\end{align*}
Subtracting this equation from the analogous equation for $y$ yields
\begin{align*}
 \sum_{i=1}^d \int_0^a (y_i z_{*}(s)-&t)^{+}\,ds
   - \sum_{i=1}^d \int_0^a (x_i z_{*}(s)-t)^{+}\,ds \\
 & = \sum_{i=1}^d \int_0^a (t-y_i z_{*}(s))^{+}\,ds
   - \sum_{i=1}^d \int_0^a (t-x_i z_{*}(s))^{+}\,ds,
\end{align*}
from which the equivalence of (\ref{eq:cmaj1}) and (\ref{eq:cmaj2})
is clear.  Thus in particular (\ref{eq:cmaj1}) holds for all $t \in (y_d z_{*}(a), y_1)$, so Step~3 is complete.

\section{Step 4}
Step~3 has provided a function $z_{*}$ that can be regarded as a continuous catalyst, as the condition (\ref{eq:step3}) is roughly analogous to the condition (\ref{eq:order_free_cm}) for a finite-dimensional vector $z$.  It is fairly straightforward to produce a suitable catalyst vector $z$ from $z_{*}$.  We provide the details in this section.

From the result of Step~3 we have that (\ref{eq:step3}) holds for all $t \in (y_d z_{*}(a), y_1)$; therefore in particular (\ref{eq:step3}) holds for all $t \in [x_d z_{*}(a), x_1]$.  This latter interval is compact,
and $\sum_{i=1}^d \int_0^a (x_i z_{*}(s)-t)^{+}\,ds$ and
$\sum_{i=1}^d \int_0^a (y_i z_{*}(s)-t)^{+}\,ds$ are continuous functions
of $t$, so we may pick $\delta>0$ so that
\begin{equation} \label{eq:sumintdiff}
  \sum_{i=1}^d \int_0^a (y_i z_{*}(s)-t)^{+}\,ds
   - \sum_{i=1}^d \int_0^a (x_i z_{*}(s)-t)^{+}\,ds \geq \delta
\end{equation}
when $t \in [x_d z_{*}(a),x_1]$.

Now let $\ell$ be an integer that is large enough to imply that
$|z_{*}(s_1)-z_{*}(s_2)| \leq \delta/2y_1 ad$ whenever $|s_1-s_2| \leq a/\ell$; such an $\ell$ must exist from the Lipschitz condition on $z_{*}(s)$.  This $\ell$ will be the dimension of our catalyst $z$.  For each $j \in \{1,\ldots, \ell \}$ pick $z_j$ as any value in the interval $[z_{*}((j/\ell)a),z_{*}(((j-1)/ \ell)a)]$.  Our catalyst $z$ is the vector $(z_1,\ldots,z_{\ell})$.

We now verify that $x \otimes z \prec y \otimes z$.
Since the sum of all components of $x \otimes z$ is equal to
the sum of all components of $y \otimes z$, we need only verify
that for all $t$,
\begin{equation} \label{eq:step4ineq}
  \sum_{i=1}^d \sum_{j=1}^{\ell} (y_i z_j-t)^{+} - 
  \sum_{i=1}^d \sum_{j=1}^{\ell} (x_i z_j-t)^{+} \geq 0.
\end{equation}

For $t \geq x_1$, the second sum is zero, so (\ref{eq:step4ineq})
must hold.  For $t \leq x_d z_{*}(a)$ we have
\[ \sum_{i=1}^d \sum_{j=1}^{\ell} (y_i z_j-t)^{+} \geq
   \sum_{i=1}^d \sum_{j=1}^{\ell} (y_i z_j-t) =
   \sum_{i=1}^d \sum_{j=1}^{\ell} (x_i z_j-t) =
   \sum_{i=1}^d \sum_{j=1}^{\ell} (x_i z_j-t)^{+}, \]
so again (\ref{eq:step4ineq}) holds.

For the remaining values of $t$, namely $t \in (x_d z_{*}(a), x_1)$, we show that the (appropriately scaled) values of the double sums in (\ref{eq:step4ineq}) are close to the values of the corresponding sum-integrals, then we use (\ref{eq:sumintdiff}) to establish (\ref{eq:step4ineq}).
We have
\begin{align*}
 \Biggl| \frac{a}{\ell} \sum_{i=1}^d \sum_{j=1}^{\ell} (x_i &z_j-t)^{+}
  - \sum_{i=1}^d \int_0^a (x_i z_{*}(s)-t)^{+}\,ds \Biggr| \\
 & = \left| \sum_{i=1}^d \sum_{j=1}^{\ell} \int_{\frac{j-1}{\ell}a}^{\frac{j}{\ell}a}
  ((x_i z_j-t)^{+} - (x_i z_{*}(s)-t)^{+})\,ds \right| \\
 & \leq \sum_{i=1}^d \sum_{j=1}^{\ell} \int_{\frac{j-1}{\ell}a}^{\frac{j}{\ell}a}
  \left| (x_i z_j-t)^{+} - (x_i z_{*}(s)-t)^{+} \right| ds \\
 & \leq \sum_{i=1}^d \sum_{j=1}^{\ell} \int_{\frac{j-1}{\ell}a}^{\frac{j}{\ell}a}
  | x_i z_j - x_i z_{*}(s)| \,ds \\
 & \leq \sum_{i=1}^d \sum_{j=1}^{\ell} \int_{\frac{j-1}{\ell}a}^{\frac{j}{\ell}a}
  \frac{x_i \delta}{2y_1 ad} \,ds \\
 & \leq \sum_{i=1}^d \sum_{j=1}^{\ell} \int_{\frac{j-1}{\ell}a}^{\frac{j}{\ell}a}
  \frac{\delta}{2ad} \,ds \\
 & = \delta/2.
\end{align*}
Similarly,
\[ \left| \frac{a}{\ell} \sum_{i=1}^d \sum_{j=1}^{\ell} (y_i z_j-t)^{+}
  - \sum_{i=1}^d \int_0^a (y_i z_{*}(s)-t)^{+}\,ds \right| \leq \delta/2. \]
Thus
\begin{align*}
 \frac{a}{\ell} \Biggl( \sum_{i=1}^d \sum_{j=1}^{\ell} (y_i z_j-t)^{+}
   &- \sum_{i=1}^d \sum_{j=1}^{\ell} (x_i z_j-t)^{+} \Biggr) \\
 & \geq -\delta + \sum_{i=1}^d \int_0^a (y_i z_{*}(s)-t)^{+} \,ds
   - \int_0^a (x_i z_{*}(s)-t)^{+} \,ds \\
 & \geq 0,
\end{align*}
so (\ref{eq:step4ineq}) holds.

Thus we have shown that (\ref{eq:step4ineq}) holds for all $t$, so
$x \otimes z \prec y \otimes z$ as desired.

\section{Discussion}  \label{sec:discussion}
A proof of Theorem~\ref{thm:main} that follows along the lines outlined here appears to be constructive enough that one could produce a crude upper bound to the minimum dimension of the catalyst, as a function of $x$ and $y$.  It may take a substantial effort to do this.  It may be of interest to determine an asymptotic growth rate of the minimum dimension of the catalyst as, say, $x$ approaches the boundary of $T(y)$.

The characterization of catalytic majorization provided by Theorem~\ref{thm:main} requires verifying a continuum of inequalities before concluding that $x \prec_T y$.  In practice, for a particular $x$ and $y$, one can generally use the properties of the $f_r$ to check all $f_r$ inequalities with a finite amount of work.  Given this, one might speculate that some of the inequalities are redundant, but we think that this is not the case: we conjecture that for any $r_0 \in \myreal$, there exists an $x$ and $y$ such that $f_r(x) < f_r(y)$ holds for all $r$ except $r = r_0$ (note that this would imply that $f_{r_0}(x) = f_{r_0}(y)$).

The methods used here suggest several possible additional mathematical investigations.  We mention one direction that seems intriguing.  The idea is that it may be fruitful to define weak catalytic majorization.  As background, we note that the usual weak majorizations (submajorization and supermajorization) can be naturally defined for infinite vectors $x$ and $y$ as follows:

An infinite vector $x=(x_1,x_2,\ldots)$ can be defined to be submajorized by an infinite vector $y$ if for all $k \geq 1$ we have $\sum_{i=1}^k x_i^{\downarrow} \leq \sum_{i=1}^k y_i^{\downarrow}$.  This definition makes sense only if for any component of $x$, there are a finite number of larger components of $x$, and similarly for $y$.  If the components of $x$ and $y$ are nonnegative and have finite sums, then this definition is equivalent to the order-free formulation that requires $\sum_{i=1}^{\infty} (x_i-t)^{+} \leq \sum_{i=1}^{\infty} (y_i-t)^{+}$ for all $t \geq 0$.
Similarly, $x$ can be defined to be supermajorized by $y$ if for all $k \geq 1$ we have $\sum_{i=1}^k x_i^{\uparrow} \leq \sum_{i=1}^k y_i^{\uparrow}$.  This definition makes sense only if for any component of $x$, there are a finite number of smaller components of $x$, and similarly for $y$.  If the components of $x$ and $y$ are nonnegative and unbounded, then this definition is equivalent to the order-free formulation that requires $\sum_{i=1}^{\infty} (t-x_i)^{+} \leq \sum_{i=1}^{\infty} (t-y_i)^{+}$ for all $t \geq 0$.

With these notions in place, we can state a possibility for weak catalytic majorization.  Suppose $x$ and $y$ are $d$-dimensional vectors with nonnegative components.  It is not assumed that the sum of the components of $x$ equals the sum of the components of $y$.  We consider infinite catalyst vectors $z = (z_1, z_2, \ldots)$ with nonnegative components (and at least one nonzero component).
A possible definition is that $x$ is catalytically submajorized by $y$ if there exists a catalyst $z$ such that $x \otimes z$ is submajorized by $y \otimes z$ and $\sum_{j=1}^{\infty} z_j^r$ is finite for all $r>0$.
Also $x$ is catalytically supermajorized by $y$ if there exists a catalyst $z$ such that $x \otimes z$ is supermajorized by $y \otimes z$ and $\sum_{j=1}^{\infty} z_j^r$ is finite for all $r<0$.

We think that if the condition on $z$ is chosen properly in both cases (we are not sure if we have it right as written), then $x$ being both catalytically submajorized and catalytically supermajorized by $y$ will imply $x \prec_T y$.

Note that with these definitions it is possible that, say, $x$ is catalytically submajorized by $y$ and the sum of the components of $x$ is equal to the sum of the components of $y$, but $x \not\prec_T y$.


On a related note, we ask: if for some $r_0 \geq 0$ we only require the sum $\sum_{j=1}^{\infty} z_j^r$ to be finite when $r>r_0$, then are the $f_r$ inequalities for $r \geq r_0$ sufficient to imply the existence of a $z$ for which $x \otimes z$ is submajorized by $y \otimes z$?  Analogously, if for some $r_0 \leq 0$ we only require the sum $\sum_{j=1}^{\infty} z_j^r$ to be finite when $r<r_0$, then are the $f_r$ inequalities for $r \leq r_0$ sufficient to imply the existence of a $z$ for which $x \otimes z$ is supermajorized by $y \otimes z$?


\ignore{
Even with an infinite number of inequalities to test, because of the ``continuity'' of the inequalities, it is a dramatic improvement over determining whether $x \prec_T y$ with a brute force search for a suitable catalyst vector $z$.  (In fact, for a given $\ell$ of moderate size, determining whether an $\ell$-dimensional $z$ exists for which $x \otimes z \prec y \otimes z$ can border on being computationally intractable.)
}

\ignore{
We hope that our main result provides some insight into the nature of the catalytic majorization relation, and perhaps also more generally into the mathematical properties of quantum entanglement.  However, our proof itself seems to provide little intuitive justification for our result.  It would undoubtedly be of interest to find an alternate (hopefully shorter) proof that illuminates the apparently fundamental nature of the $f_r$ family of inequalities.
}

\appendix*
\section{}
Here we prove a claim that was used in Section~\ref{sec:sufficiency_outline} to show that in our sufficiency proof we need only consider those $y$ whose components are all nonzero.  We restate the claim here.
\begin{claimnn}
Suppose $x$ and $y$ are $d$-dimensional probability vectors with components in decreasing order, and such that $x_1 < y_1$ and $x_d > y_d$ and for all $r \in \myreal$ the inequality $f_r(x) < f_r(y)$ holds.  Then there exists a $d$-dimensional probability vector $y'$ whose components are all nonzero, with $x_1 < y'_1$ and $x_d > y'_d$ (we assume the components of $y'$ are in decreasing order), such that $f_r(x) < f_r(y')$ for all $r \in \myreal$, and such that $y' \prec y$.
\end{claimnn}

\emph{Proof.}
If all components of $y$ are nonzero, we can simply take $y'=y$, and we are done.  Therefore we suppose $k$ components of $y$ are $0$, where $k \geq 1$.  Let $w$ be the $d$-dimensional vector for which the first $d-k$ components are $0$, and the remaining $k$ components are $1$.  For $n \geq 1$ let $y^{(n)} = \frac{n-1}{n}y + \frac{1}{nk} w$.  Clearly, for sufficiently large $n$, say $n \geq N_0$, the components of $y^{(n)}$ are in decreasing order and $y^{(n)} \prec y$.  Note also that $y^{(n)} \prec y^{(n+1)}$ for $n \geq N_0$.

Choose $N_1$ to be large enough that $1/N_1 k < x_d$.  Choose $r_1<0$ so that if $r<r_1$ then
\[  \left(\frac{1}{N_1 k} \right)^r > x_d^r d.  \]
Choose $N_2$ to be large enough that $\frac{N_2-1}{N_2}y_1 > x_1$.  Choose $r_2>1$ so that if $r>r_2$ then
\[  \left(\frac{N_2-1}{N_2} y_1 \right)^r > x_1^r d.  \]

For general $x$, $y$, and $r$ we define
\[ F(x,y,r) =
  \begin{cases}
    \frac{1}{r(r-1)}( \ln \sum_{i=1}^d y_i^r - \ln \sum_{i=1}^d x_i^r),
       & \text{if $r \neq 0,1$;} \\
    -\frac{1}{d} ( \sum_{i=1}^d \ln y_i - \sum_{i=1}^d \ln x_i ),
       & \text{if $r=0$;} \\
    \sum_{i=1}^d y_i \ln y_i - \sum_{i=1}^d x_i \ln x_i, & \text{if $r=1$.}
  \end{cases} \]
If all components of $x$ are nonzero then $F(x,y,r) > 0$ is equivalent to $f_r(x) < f_r(y)$.  It is straightforward to verify that for fixed $x$ and $y$, both with all components nonzero, $F(x,y,r)$ is continuous in $r$ over all of $\myreal$.  For fixed $x$ and $r$ the function $F(x,y,r)$ is Schur-convex in $y$.

For our specific $x$ and $y$ (with $y^{(n)}$ constructed from $y$) let $g_n(r) = F(x,y^{(n)},r)$.  We then have that for all $n \geq N_0$ the function $g_n(r)$ is continuous in $r$.  Also if $n \geq N_0$ then $y^{(n)} \prec y^{(n+1)}$ and so $g_n(r) \leq g_{n+1}(r)$ for all $r \in \myreal$.  Finally, for all $r \in \myreal$ it can be verified that $\lim_{n \rightarrow \infty} g_n(r) > 0$ (specifically, the limit is $F(x,y,r)$, and in particular it is $+\infty$ when $r \leq 0$).

From these properties of $g_n(r)$, a subclaim that we state below implies that there exists an $N$ such that for all $r \in [r_1,r_2]$ we have $g_n(r)>0$ when $n \geq N$.

Let $y' = y^{(n)}$ for $n = \max(N_0,N_1,N_2,N)$.
Now if $r \in [r_1,r_2]$ then $g_n(r)>0$ so that $f_r(x) < f_r(y')$.
If $r<r_1$ then we have
\[ \sum_{i=1}^d x_i^r \leq x_d^r d < \left(\frac{1}{N_1 k}\right)^r
   \leq \left(\frac{1}{n k}\right)^r \leq \sum_{i=1}^d (y'_i)^r, \]
which implies $f_r(x) < f_r(y')$.
If $r>r_2$ then we have
\[ \sum_{i=1}^d x_i^r \leq x_1^r d < \left(\frac{N_2-1}{N_2}y_1 \right)^r
   \leq \left(\frac{n-1}{n}y_1 \right)^r \leq \sum_{i=1}^d (y'_i)^r, \]
and again we have $f_r(x) < f_r(y')$.
Thus $f_r(x) < f_r(y')$ for all $r \in \myreal$, so the proof is complete except for establishing the subclaim.

\begin{subclaimnn}
Suppose $[a,b]$ is an arbitrary closed interval, and $g_n(r)$ satisfies the following:
\begin{itemize}
\item[(i)] for all $n \geq N_0$ the function $g_n(r)$ is continuous in $r$ over all of $\myreal$;
\item[(ii)] $g_n(r)$ is increasing in $n$ when $n \geq N_0$ and $r \in \myreal$; and
\item[(iii)] $\lim_{n \rightarrow \infty} g_n(r) > 0$ for all $r \in \myreal$.
\end{itemize}
Then there exists an $N$ such that $g_n(r)>0$ for all $r \in [a,b]$ and $n \geq N$.
\end{subclaimnn}

\emph{Proof of subclaim.}  For all $n \geq N_0$ let $E_n = \{\, r \mid g_n(r)>0 \,\}$.  Then from (i) each $E_n$ is an open set, and it follows from (ii) that $E_n \subset E_{n+1}$.  From (iii) we have $\bigcup_{n \geq N_0} E_n = \myreal$, so in particular $[a,b] \subset \bigcup_{n \geq N_0} E_n$.

Because the interval $[a,b]$ is compact, there exists a finite subset of $\{E_n\}_{n \geq N_0}$ that covers $[a,b]$.  In view of the fact that $E_n \subset E_{n+1}$, there is thus an $N$ for which $[a,b] \subset E_N$, which implies that $[a,b] \subset E_n$ for $n \geq N$.  Thus the claim holds using this choice of $N$.
\myqed

\emph{Remarks.} Our subclaim appears to be a simpler relative of Dini's theorem, and our proof of the subclaim is similar to a standard proof of Dini's theorem.  The fact that when $n$ is large enough $y^{(n)}$ has the desired properties can also be proved without the subclaim using the definition of the $f_r$ directly.


\begin{acknowledgments}
The author is indebted to Sumit Daftuar and Michael Nielsen
for introducing him to (what we now call) catalytic majorization, and for useful discussions and encouragement when he started investigating the subject.
\end{acknowledgments}

\bibliography{maj2007}

\end{document}